# Renewing computing paradigms for more efficient parallelization of single-threads

János VÉGH

*University of Miskolc, Hungary*

**Abstract.** Computing is still based on the 70-years old paradigms introduced by von Neumann. The need for more performant, comfortable and safe computing forced to develop and utilize several tricks both in hardware and software. Till now technology enabled to increase performance without changing the basic computing paradigms. The recent stalling of single-threaded computing performance, however, requires to redesign computing to be able to provide the expected performance. To do so, the computing paradigms themselves must be scrutinized. The limitations caused by the too restrictive interpretation of the computing paradigms are demonstrated, an extended computing paradigm introduced, ideas about changing elements of the computing stack suggested, some implementation details of both hardware and software discussed. The resulting new computing stack offers considerably higher computing throughput, simplified hardware architecture, drastically improved real-time behavior and in general, simplified and more efficient computing stack.

## 1. Introduction

Despite of the bright and sound successes of computing on practically every single field of life, future development of computing is in serious danger [1]. After approaching and reaching the technological bounds, the only hope to increase further computing performance is parallelization. In todays computing, a plethora of parallelization principles and technologies is available and in use [2]. Unfortunately, the efficacy of computing gets the worse the more efforts are expended in parallelization. In HW, the parallelization resulted in architecture of frightening complexity [3], limiting the clock speed [4], etc. The first warning signs about reaching a dead-end street triggered introducing multi- and many-core processors (MCP) as a prospective way of development, although it was known [5] that their performance is seriously limited. For now even that direction was declared as broken [6] and the age of "Dark Silicon" [7] entered.

Parallelization is not much more successful in the SW world, too. Although running very similar independent calculations in several independent processors "in parallel" is possible, and the supercomputers are highly successful, the real-life tasks show variable degree of parallelization [8]. Because of this, parallelizing a general single-threaded task cannot be solved effectively on traditional architectures (and thinking traditionally), even if special HW solutions are used to accelerate the task using sequential/parallel sections [9].

The way to more performant parallel systems leads through more performant single-threaded processors, so ways to increase single-threaded performance are desperately researched. It became clear that further development in computing is not possible without reinterpreting the computing paradigms themselves. Neumann was aware of this need: *"After the conclusions of the preliminary discussion the elements will have to be reconsidered"* [10]. Below, some paradigms are scrutinized, mainly from the point of view of possible parallelization.

Parallelism has different meaning in SW and HW worlds, although in both cases the main goal is to use more computing units at once. In HW, since the beginnings of computing, the computing is implemented as a simple sequential one-dimensional graph (although with repetitions due to loops and branching). The processor considers the program as a one-dimensional stream and deals with the instructions one-after-the-other. Real performance increase can only be achieved through "cheating", when the processor considers not only the current instruction, but also its (possible) followers (like out-of-order, speculative, etc. evaluation). To provide extra computing power, of course extra (complete or partial) computing units are needed, which always remain hidden: the processor must persist the view as being the only computing component, as required by the one-to-one correspondence between processor and process. Actually it means that the only way of increasing performance through this hidden parallelism is to make the one-dimensional graph "thicker".

The SW world is in direct connection with the real world: experiences the needs like running different processes "at the same time" as well as interacting with and modeling of working environment running in parallel regime. A special software layer (called operating system, OS) between the parallel world and the single-thread processor must provide the proper illusion towards both parties. The single processor believes it is running a single process (which is always true, although the process changes frequently), and all processes believe they have their own processor (although for just a fraction of time). In this sense parallelization results in a two-dimensional graph, comprising several one-dimensional (maybe "thicker") graphs. Unfortunately, because of utilizing shared resources, the one-dimensional graphs must be synchronized, which action has its considerable expenses [11].

In summary, the two main obstacles on the road towards a more performant computing are the final speed of the light (forcing the miniaturization of electronic components and thus leading to the "thermal wall") and the too restrictive paradigm interpretation, that *the same processor* must be present in all process-processor relations and *the complete duration of the lifetime of a process.* There is no chance to alter the first, but one can try to change the second one. In the followings some of the bad practices are pinpointed in section 2 and the idea of a new computing paradigm called Explicitly Many-Processor Approach (EMPA) is introduced in section 3. Some ideas about its implementation as well as its consequences are discussed in section 4. The EMPA-related developments, including the tools, are presented in section 5, together with some of the variety of different solutions using EMPA. Some performance consequences are also highlighted there.

**2. Limitations of traditional solutions**

Till now, computing was successful in using the rigid interpretation of the process-processor correspondence. Unfortunately, the components of the computing stack, from compilers to buses, from processor to memories, are built in the same approach, which reflect the technical level of the age 70 years ago.

*2.1. Rigid architecture: theoretical parallelism*

Architecture of a computer is traditionally inflexible, unlike the tasks to be solved. The example discussed here is borrowed from [2]. The task is to perform the simple calculations shown in the inset on the left side of Fig. 1. On an ideal processor, one could load all needed arguments in the $1^{st}$ clock cycle, make the multiplications in the $2^{nd}$ and calculate the two results in the $3^{rd}$ one. To do so, one would need 4 loaders and 2 arithmetic units, or 4 simple complete processors. In a real processor with rigid architecture, one needs 4 complete processors, which must be present all the time, although in the last two cycles only 2 of them are utilized. The other two are also heating, however.

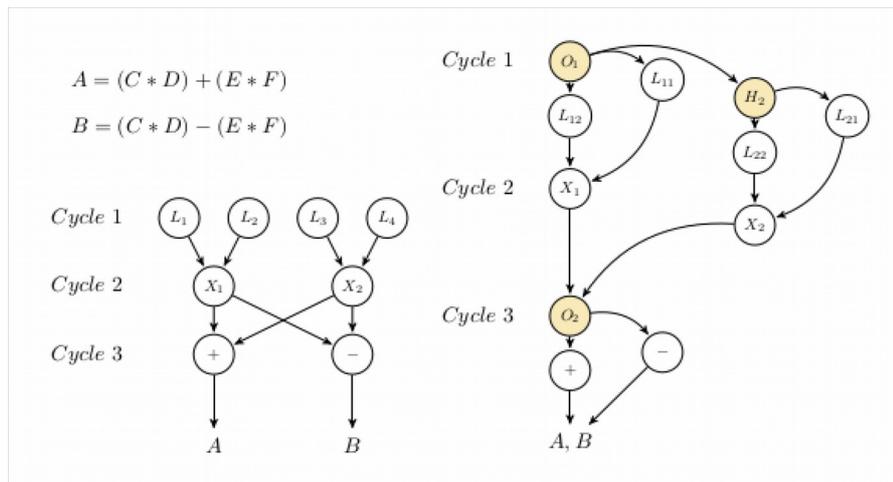

*Figure 1: The theoretical parallelism example*

In a real, but flexible implementation, see the right side of Fig. 1, it is possible to rent some processors from a central pool when the task requires so. This action happens in the shaded nodes. The rent processors make only one operation, then they are put back to the central pool and are available for performing another task. As displayed, this flexible processor can solve the task in 3 clock cycles, although some extra time is needed for renting and returning a new core, and also more capable memory (like [12]) shall be used.

The presently existing inflexible architectures (two of them are displayed in Fig. 2) can only have a fixed number of extra processing units. The left side shows a dual-issue processor: it has a unit for loading and another one for calculating. The two units can run in parallel. Since in the example task there are 4 tasks for both units, one may even hope to achieve a speedup factor of 2. The data dependence and the components built in Single Processor Approach (SPA), however, limit their utilization: no calculation can be performed until both operands are loaded and only one operand can be loaded at a time using the components build with SPA in mind. Because of this, it is only Cycle 3 when both units can work. As a consequence, the 8 operations can be done in 7 cycles, which is a moderate gain for investing a double hardware cost. Note that using dual-channel memory (like [12]) would enhances its operation.

The right side of the figure represents another inflexible architecture. The dual-core processor seems to be a better option: the two cores can perform the two operations independently, so the first few steps of calculations can run really in parallel (although the operands can only be loaded one after the other when using SPA components). The real problem appears at cycle 4: how to access the data belonging to the other (independent) processor. One solution can be to store and load the data for and from the other party: it requires 4 extra memory cycles, and all issues of the shared memory. In exchange, the task is solved in 6 cycles, not much more efficient, than the double issue architecture.

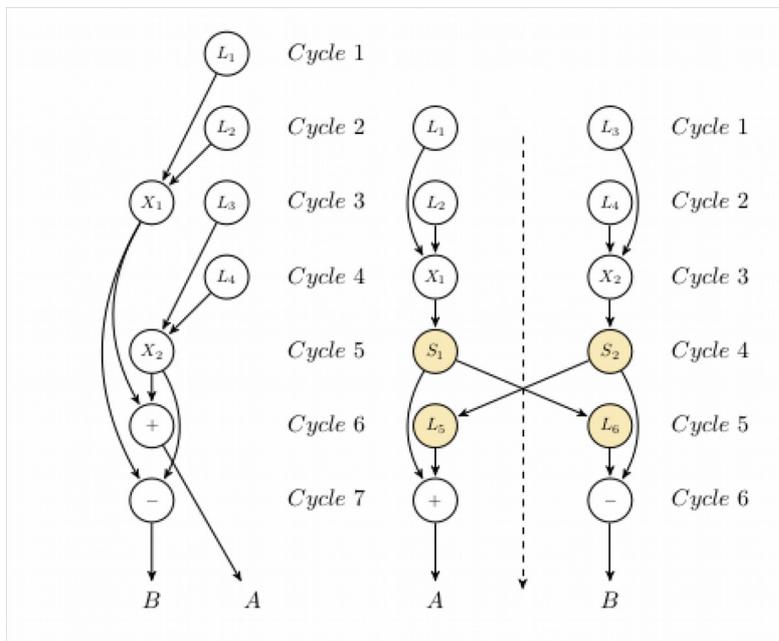

*Figure 2: Dual-issue and dual-core implementations of the example calculation*

In summary, a general calculation task cannot be solved efficiently with a rigid architecture [13]. Typically most units are idle in most part of the execution time, and the speedup is not proportional at all with the invested hardware. More (theoretically obsolete) states must be inserted into the execution graph, and the theoretically possible parallelism cannot even be approached with such architectures. The dynamic architecture displayed in the right side of Fig. 1 appears promising, but surely cannot be implemented using conventional components and thinking, all based on SPA.

*2.2. Single-processor approach: supercomputer development*

Today the inexpensive high-performance processors are ubiquitous, dozens or even hundreds of independent processors (called cores) are present in a single die. However, all these cores are CPUs (*Central* Processing Units), which clearly shows they are built in SPA. As Amdahl pointed out 50 years ago, *"the organization of a single computer has reached its limits and that truly significant advances can be made only by interconnection of a multiplicity of computers in such a manner as to permit cooperative solution."* [14] Despite this prophecy, all computer systems, including supercomputers, are built even today in SPA. The most obvious proof of this is as the supercomputer performance developed in time, see Fig. 3.

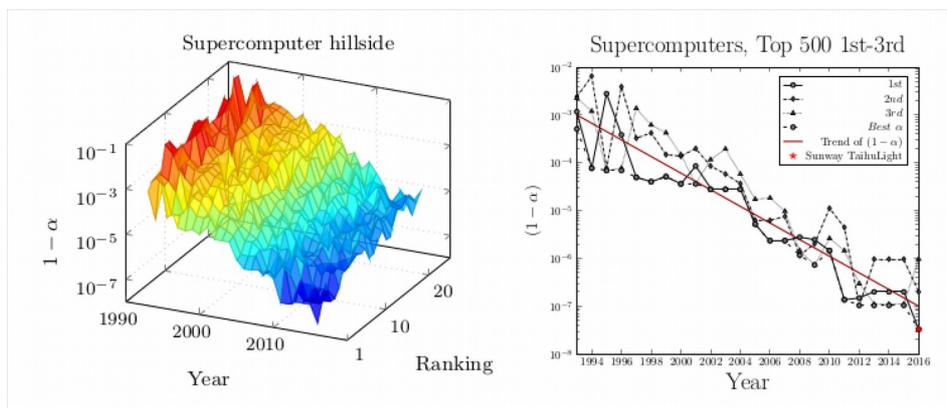

*Figure 3: Timeline of development of the non-parallelizable fraction in the case of supercomputers*

The supercomputer development is an excellently documented [15] story of computing. Performance data are available for its complete 24-year history, independently of processor type, architectural solution, manufacturer, etc. Although Amdahl wanted to draw the attention to the *limitations of performance increasing* when using SPA, his followers constructed the so called Amdahl's law, which is generally believed to be valid only for software activities.

As discussed in [16], Amdahl's idea shall be interpreted that quantities are referring to *execution time* rather than the *number of executed instructions*. In this interpretation Amdahl's law is valid for parallelizing any kind of time-related activities, including solving tasks using complex HW/SW systems. From the published data one can conclude the fragment of time spent with non-parallelizable fraction (denoted by (1-α) in Fig. 3) of the benchmark task used to qualify the computer system. This parameter strongly decreases with the year of implementation at a given rank, and slightly increases with the ranking of the computer in a given year (see the left side of Fig. 3), without considering processor type, manufacturer, technology, etc.

To establish a more quantitative statement, some of the data from the performance hillside are displayed on the right side of Figure 3. As shown, the (1-α) parameter of the first three (by performance parameter $R_{max}$) supercomputers follow a trend line (just drawn to guide the eye), essentially the manifestation of Amdahl's law, seems to be another new "*Exponential law of computing growth*" [17]. It is exactly what Amdahl contended a quarter century before the beginning of the supercomputer age: "*the effort expended on achieving high parallel processing rates is wasted unless it is accompanied by achievements in sequential processing rates of very nearly the same magnitude*" [14].

*2.3. Improper computing stack*

The efficacy of computing with general-purpose processors (using SW layer(s)) is tragically low compared with solutions in ASIC [18]. When accommodating a general purpose processor to a very specific problem, typically at least 3-layer computing stack is utilized: processor, OS and application. For the OS developers the processor was a "black box", and similarly the OS for the application developers. Some extremely necessary functionality appeared in the next releases of the HW and OS, but most of these functionalities are implemented in OS or application level in a suboptimal way.

A nice example showing numerical values of performance loss is presented in [19]. A simple (and frequently used for controlling shared resources in modern applications) facility is the mutex (as a kind of simplest semaphore). Its basic functionality is just to handle 1 single bit, which task can be solved in one single clock cycle. Such functionality is not provided by the HW (the semaphores/mutexes provided in HW are designed for inter-processor communication, rather than inter-task communication). If a mutex is implemented in SW and uses the OS facilities, it consumes cca. 2700 clock cycles, even if no context change used. As described in [20] [21], the context change in modern operating systems takes time in the order of $10^4$ clock cycles. This is one of the reasons why very task specific (mostly reconfigurable) solutions outperform solutions using general purpose processors in the order of $10^5$-$10^6$ times, even when driven by at least an order of magnitude lower clock speed. When one could reduce those losses through using a more reasonably assembled computing stack, a strong performance increase could be achieved. This example suggests that *it is worth to scrutinize whether the computing stack is properly layered*.

*2.4. Missing features*

The computer is a single-purpose device, it can only compute. As "*if all you have is a hammer, everything looks like a nail*", it computes addressing, termination and branching conditions, etc. information, not speaking about features missing for making multiprocessing effectively [22]. For example, when summing up elements of a vector, the only payload operation is adding a new term to the previous sum. However, the new term must be addressed, the number of terms counted, the termination condition computed, and a conditional jump is used to repeat the same calculation several times. In this simple case some external HW (like another, cooperating core) could perform those helper operations in parallel with the payload operation, if it could receive the data for the calculation and return the result. *Adding the ability to cores to communicate with other cores with this goal, one can reduce the number of instructions to a fragment of the original number*, and in this way apparently increase the performance by a factor 2-4 in this operation, at the price of using one more processing unit.

The general purpose processors dealing with general purpose computations cannot adapt their architecture to the task and so in the general case [13] most of their computing capability remains idle. The Digital Signal Processors demonstrate, that contracting some arithmetic operations pays off in terms of performance. In DSPs, the high performance instructions are implemented as inflexible HW subcomponent, and can only be utilized if the actual task enables it, otherwise that capable and expensive HW remains idle. In a flexible architecture having cooperating processors, 'ad-hoc' computing assemblies can be organized for the time of performing such types of computing.

In the previous example, one can notice that the intermediate sum is a do-not-care intermediate result. And, the need of using the intermediate sum and the inflexible computational atomic unit 'machine instruction' is the reason, why this kind of summing cannot be parallelized: one of the operands is the temporary sum and the machine instruction reads it from a temporary register before making the addition, and writes it back when exiting the machine instruction, just to enable to read it again in the next iteration of the loop. If the control unit would be able to delegate the task of calculating the individual terms to another cores, and allocate an adder for adding the returned calculated summands on the fly, even this classical not parallelizable task could be parallelized, at the price of using several computing units. Obviously, for that goal also the control unit must be "programmable" (or better: configurable), clearly indicating the need of 2-level programming.

*2.5. Predictability of operations*

Although lack of predictability of execution time of an instruction is a consequence of the missing features [22] of the processor, its importance and bad consequences deserve a separated section. The SPA resulted in using *interrupt* for sharing the only available processor between multiple tasks. Unfortunately, this approach is used even in the presence of several processors (cores). In the latter case any of the processors could serve the external demand, but the running process is interrupted and its processor is taken to service the external demand. As mentioned, it is a very expensive operation in terms of computing time, mainly because of the necessary (both HW and SW) context change, but its worse effect that it makes the real-time execution of adjacent machine instructions unpredictable. Although special solutions like [23] are used successfully for low core numbers in solving even hard-real-time problems, using conventional SPA architecture does not enable developing general methods of using benefits of many-core processors.

In tasks sensitive to the real execution time (and it is getting more and more emphasized with the spread of complex cyber-physical systems) special care must be exercised with scheduling, for example to avoid issues like priority inversion or stack overflow. This task, again, remains for SW. However, when mitigating the effect of the complex SW operation, the SW is getting even more complex, and the real execution time not only longer, but also less predictable. If the control unit would be able to use dedicated processor for servicing external request, one could economize the time needed for context changing, and in addition the processor could answer the requests with ideally low latency and service times.

**3. The renewed paradigm**

Amdahl [14] wanted to draw the attention also to the fact that the *Single Processor Approach* has serious limitations when assembling many-processor systems. As demonstrated above, the present many-processor systems are assembled from segregated rather than cooperating processors. In addition to that, many-processor systems are assembled from components manufactured for single-processor systems by engineers trained for single-processor systems, the main reason is the requirement of compatibility: all of the programming infrastructure, including even most programming languages, is built on the idea of sequential execution [24].

Since Arvind and Iannucci [22] it is known that some features are missing from the processors, which would enable them to handle several processes effectively. The lack of those features led to implement the missing in HW features in SW, with minimum HW support; just to keep the illusion that no more processes exist. The computer can really only *compute;* even those operations, which result in providing only signals like exceeding the requested number of operations, are provided with making computations, using the complete computing infrastructure. Even when placing the conventional inflexible architectures side-by-side, as in the case of many-core processors, they cannot help each other effectively.

As a common background of those obstacles one can recognize *the paradigm about the one-to-one correspondence between processor and process.* To change the paradigm, however, other outstanding issues, concluded from the utilization requirements and experiences, must be considered.

*3.1. The need for renewing paradigm*

The really big idea of von Neumann about computing was to introduce an interface between mathematics an engineering, rather than inventing some kind of architecture. After accepting that interface, mathematical research assumes the behavior of technical computers to satisfy those requirements, and also the engineers must consider those constraints on their designs. During development, however, both the engineering possibilities allowed utilizing more and different solutions, and the utilization mode forced to apply the same computer for different goals. However, to keep the interface unchanged, HW engineers provide the illusion that in their designs only one processing unit provides the computing performance, and the SW engineers introduced the interface OS which provides the illusion for the tasks that they have their own processor. Of course, both parties know they are cheating: HW provides the possibility to *interrupt* the running process (why if there are no other processes) and SW provides tools to make exclusive use of some resources (why if there are no other processes to use the resource). Despite this *implicitly multi-process working regime*, computing faces no unsolvable problems, although its performance is drastically lower than it could be.

*3.2. Minor changes to interpreting the paradigm*

Fortunately, von Neumann formulated the requirement for the order of instruction execution that the task of the control unit is to provide a "*proper sequencing*" [10] of executing machine instructions, which allows to utilize several "tricks".

On the HW side, initially it was interpreted that the processor cannot receive new instruction until the current instruction is completely executed. However, the "need for speed" forced introducing interrupts for handling unexpected events in a reasonable way. The control unit could provide the expected "proper sequencing", although it became considerably more complex. Similarly, when the machine instruction execution was separated to different stages and it was recognized that the electronic circuits were actually used only in a fragment of the total execution time, the pipelining introduced separate signals for "instruction execution finished" and "ready to accept new instruction". This did not break the requirement of "proper sequencing", but the requirement to consider the data and control dependence required incomparably complex operation (like register renaming) of the control unit.

Even using Very Large Instruction Words, Out-of-Order evaluation, etc. could be accepted at the price of making the *changed sequencing* invisible for the external world. Branch prediction and speculative evaluation implement complete alternative calculation processes, without breaking the principle, but at the price of introducing frightening complexity in the control unit. The parallelization of instruction execution within the processor has the potential for increasing the performance by hundreds of times [25]. However, because *without external help* the processors cannot use wide instruction window, the actual parallelism rarely exceeds the value 3-5 as found early [26] . Despite this, "*instruction-level parallel processing has established itself as the only viable approach for achieving higher performance without major changes to software*" [4].

On the SW side, introducing the most successful HW accelerators: register file, the SW unit "thread" and the multi-tasking OS with its operating modes caused major problems. The processors could work with a considerably lower number of internal registers [27]. Although the atomic unit of execution itself, the machine instruction, does not know about any context, the introduced register file (and later local cache) cause some unwanted bias between adjacent instructions. As a consequence, the processes have some HW context, and with introducing multi-tasking a SW context also appeared. Even in the case of the simplest multitasking (interrupt handling without OS) at least (part of) the HW context must be saved and restored. In modern multi-tasking environments with considerable complexity, due mainly to changing operating mode and SW context, using a single processor to different goals became extremely expensive: a context change may require clock cycles in the order of $10^4$ clock cycles [20] [21] and introduces a considerable load on memory traffic. Again, these extra operating modes do not break the paradigm, but considerably contribute to the overall low efficacy of general purpose computing systems [28] [21].

*3.3. A major change to interpreting the paradigm*

The first step towards changing the paradigm is to revert the direction of connection between processor and process. By convention, in SPA only one processor exists, and *the task is to provide a process for the only available processor*. In the today's multi-tasking world several processes and several processors exist, and *the task is to find an available processor for a runnable process*. As the end result, the one-to-one correspondence between a processor and a process persists, the difference is that *not the same processor appears in all relations*. The approach is called Explicitly Many-Processor Approach (EMPA) [16].

One can notice that in EMPA essentially *part of the duties of the OS is taken over by the processor*. To handle the processes (without the help of the OS) the processor must have the abilities conventionally implemented in the OS and also the SW must be able to provide unusual functionality, like suspending its operation, again an OS functionality. Because of these, a new execution unit Quasi-Thread (QT) is introduced. It is derived from HW machine instruction and SW thread, and inherits best features of both. Its size is between one machine instruction and a complete SW thread, and its remarkable feature is that it may comprise another QTs, to arbitrary depth. QTs attempt to be independent as much as possible: they comprise code chunks which are completely provided with the necessary SW context and return only a minimum amount of result. Here a trade-off between using *less quick-access registers* and using *more computing resources* must be made.

Because processors may be able both to send and receive signals for handling QTs, as well as sending and receiving (limited amount of) data in a synchronized way, processors may cooperate and help each other in solving a task. A core and a QT it runs have a much closer cooperation than processor and process in conventional computing: they are used in an interchangeable way during the lifetime of a QT, depending on whether HW or SW aspect receives more focus in the actual statement. The offset of the code chunk the core runs and the physical ID of the core identify the QT uniquely, enabling unique mapping between compile-time code addresses and run-time physical cores running that code.

**4. Ideas for implementation**

The ideas outlined above do not prognose an easy path to implementing them. Fortunately, most of the methods of implementation are already available either in HW or SW fields, or as a solution in reconfigurable (RC) technology. Some of them should be adapted to this unconventional architecture. It is sure that the accelerators designed for SPA do not work at all or work with lower efficiency for EMPA, but surely other (slightly different) accelerators will boost the operation of EMPA processors, too.

*4.1. A platform for implementing the new approach: the multi-core processors*

A decade ago, the processor technology reached a turning point. The complexity of processors cannot be increased any more in a reasonable way (even, further increase would cause the decrease of the clock speed [4]). The underlying technology is, however, still able to deliver more transistors on the same die, so the era of producing multi-core and later many-core processors (MCP) started. Those cores are not more, than several segregated (and usually more or less simplified) processors, and the computing utilizing MCPs inherited all issues connected to SPA. It was quickly realized that the performance increase is far from being linear [5] and finally manufacturing MCPs (mainly because being manufactured in SPA) has been declared as broken [6].

In MCPs the cores are located in close proximity to each other, thus allowing unusual operations, like core spilling [30]. This special kind of load balancing directs (parts of) threads with complete data control to another cores and results in dozens of percentages of performance increase. This technology suggests the idea to *share the job between nearby cores*, in a somewhat more predefined way, and calls the attention to the fact that *processing unit is one of the needed resources*. To share a job, the control unit must be able to re-delegate part of the job originally delegated to the core to some another core. Since the QTs may embed another QTs, and the QTs can be made (quasi) independent of each other, the task for a core is to recognize that execution of an embedded QT follows and to be able to signal this state to the processor-level control unit (supervisor). The supervisor must be able to receive that signal and (depending on the actual HW situation) either deny the request or to find a free core and to establish the necessary (data and control) connection between the two cores. The core that originally received the task for execution remains responsible for execution; awaits termination of the outsourced job and receives its result (all this in a transparent way), and in the meantime can work in parallel with the core "rented" in this way.

Cores are kept in a pool under control of supervisor in "power economy" mode, using two sleep transistors [31]. Upon request from the control unit, a core is revived and after use returned to the pool and can sleep again. Notice that this suggested work regime fulfills the requirement of "proper sequencing" and that it is essentially a reincarnation of the OS thread handling policy with two important differences. Once, it happens at processor level and HW speed and second that in a true parallel way unlike the apparently parallel way in the OSs. The processor should not worry about breaking dependency: it was carefully checked by the compiler and hints for implementing the dependency were transmitted in the executable code.

The conventional inflexible architectures are not suitable for such regime of operation. The "desktop supercomputer" [9] and similar constructs fail because they implement master-slave relationship between the cores. Instead, a parent-child relationship (of several generations) should be used: a parent can "rent" any number of children, but a child can have only one parent, and the parent-child relationship persists only in the time intervals the cores need to cooperate. The control unit must handle the problem of mapping virtually infinite number of QTs to the finite number of available cores. Since QTs have SW-thread like features, starting a new QT will block the execution of the requesting QT until resources get free again.

Although the cores work in a *coordinated* rather than independent regime, and so the cores can perform calculations independently, child cores may return their result not exactly at the time when their parent core needs it. To avoid synchronization issues, latches are used to store the returned data temporarily: a child writes its result to a latch when its thread terminates, and its parent waits for the termination of the child, then reads the result from the latch.

As was pointed out [25], the Instruction Level Parallelism can achieve parallelism of level several hundreds, if the instruction window is wide enough. The processor cannot solve this task alone: it works in real time, with limited storage resources. The compiler has enough time and resources to discover all possible parallelism, but does not know the actual HW stage of the processor, and also has no way to transmit its discoveries to the processor. The two players together, however, can provide excellent solution: the compiler provides hints about all possible cases and the processor needs to check the actual HW situation and chooses the proper hint. Of course, the information on parallelization must be inserted into the executable code (hints for the control unit, inserted into the stream of executable codes), and cores must recognize (during their pre-fetch cycle) the code for the supervisor, and to organize the execution jointly. This mechanism is much similar to that of the coprocessors.

### 4.2. Implementing cooperation among cores

Mainly to maintain compatibility with th conventional computing, cooperation among cores must be formulated using the conventional terms of computing. Such a term is the scratchpad register. Conventionally, it is a special (short access time) storage area, with a well defined access address. It is not unusual to use special-purpose registers, which have a register address, but they provide special functionality (like program counter or stack pointer). A similar construct can be used to enable inter-core data and control communication. These pseudo-registers have well defined access addresses (as if they were an item in the register file), but some special functionality is attached to them, and their operation is controlled by the supervisor. The cores believe they are reading/writing their own register (they use a register access code in their own scope), but the attached functionality allows to provide situation-dependent functionality: the content of another register in another core or result of independent functionality, depending on the actual mode of operation. The control unit behaves as a proxy in transferring data and control signals from one core to another.

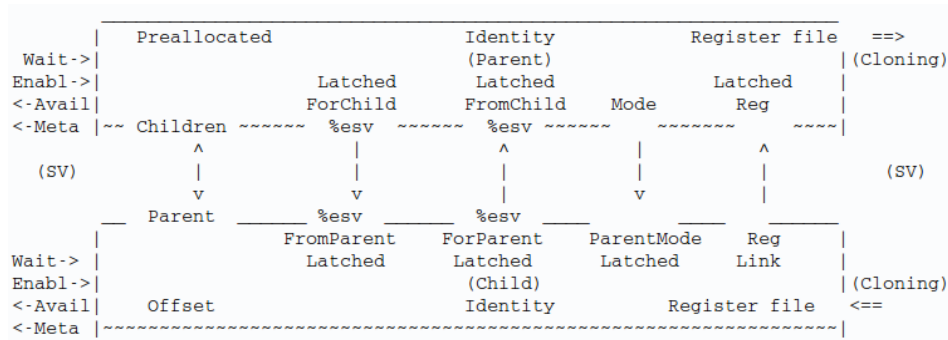

*Figure 4: The scheme of control and communication among cores in different roles in the parent-child relationship of EMPA*

The resulting control signals and extra functionality is illustrated in Fig. 4. The top core is in role "Parent", and the bottom core is a "Child". The cores can notify supervisor when they find a meta-instruction, and can block the QT running on the core by enabling/disabling it, really providing thread-like behavior. As shown the cores' pseudo-register is mapped to different physical registers, which communicate with each other. Register file and link registers are cloned between the paricipating cores using core spilling [30]. The functionality of the control unit trivially separates to two layers. The control unit of the individual cores is slightly extended, and new functionalities belong to a second, processor-level layer (called supervisor, SV). A proof-of-concept implementation in C++ has been implemented, for details see Section 5.

For the implementation, ideas of FPGA architectural principles can be borrowed. To enable quicker communication a lot of extra wiring is needed from the cores to the control unit, and -similarly to the block RAMs in FPGAs- some latches must be placed next to the cores, serving as Inter-Core Communication Blocks. Physically, these blocks correspond to the pseudo and latch registers, and their access time can be between the register ($L_0$) access time and $L_1$-cache access time. They may also contain some extra HW for non-conventional operation. Essential that the run-time re-configuration needs no routing, so the architecture can quickly adapt itself to the task.

*4.3. Programming the unconventional architecture*

Code transmitted to the processor comprises conventional code for the cores and supervising code for the control unit. Inter-core communication can be carried out trough (apparently) using registers and also some special instructions are needed for synchronizing operations as well as operating some special regimes to take advantage of the EMPA architecture. For programming, an extension of the Y86 assembler is used. New meta-instructions generate code intended for programming (or rather: configuring) the supervisor.

First of all, one needs instructions for segmenting the code, i.e. cutting it into QTs. The instructions `QCreate` and `QTerm` delimit the code chunk, corresponding to creating and terminating a QT. Notice that the latter can also be used to return control voluntarily to the initiator. The syntax

```
LabelC: QCreate   LabelT, %reg
           <body of QT>
LabelT: QTerm
```

allows delimiting the QT and specifying the register for returning result. Upon creating a QT, the complete register file is cloned into the register file of the rented core and upon terminating, executing `QTerm` causes returning the content of the named register to the corresponding register in the parent core. One can notice that this kind of operation is quite similar to 'fork' in handling thread, with the essential difference that for the operation a new computing resource is also provided and that it occurs at hardware level, without expensive context changing.

As discussed, parent core remains responsible for executing the process it received, and can only terminate if all its children already terminated. Parent QT can issue `QWait` instructions to its SV, either to wait for a specific QT (specified with its code offset address) or all of its children. The scope of the wait instruction can be the children of the QT or its sisters (the other children of QT's parent). Issuing the `QWait` instruction also means that the parent is ready to receive the register content from the child (stored in a latch register by SV) to its own register, so no synchronization issues can happen. After executing `QCreate`, parent core considers the QT as *logically* already executed, and can make sure that it was *physically* executed when `QWait` tells so.

In addition to the conventional registers, EMPA pseudo-registers `%esv, %ecc` and `%eno` have also been introduced for inter-core communication. Register `%esv` has several, context-dependent functionalities, `%ecc` is used when the condition codes are to be returned and `%eno` when syntax requires the presence of a register argument, but the corresponding functionality is not required. The detailed discussion of the context, utilization, examples, etc. can be found in [36].

*4.4. Cooperation in performing calculations*

The meta-instructions above enable the cores to cooperate akin threads of OS do (although by orders of magnitudes quicker [11]). The possibilities, however, are much broader. When considering a loop, control unit can take control back to the beginning of the loop the requested number of times, without using non-payload calculations or jumping instruction(s), reducing in this way the number of executed instructions. Since the computational density [32] cannot be increased any more, ad-hoc computing accelerator assemblies can be organized, which – in some sense following the idea of pipelining – can assist in preparing the next iteration while parent core is busy with performing the main calculation. The flexible size of QTs allows to parallelize complete ranges of computation layers, from thread-level to instruction level. When changing to EMPA (i.e. explicitly considering, that several processing units having their own register files are available), the compilers can also discover much higher level of parallelism.

To enable using more than one cores, parent core must allocate them in advance, in order to be sure the concurrent QTs will not occupy them. In EMPA, meta-instruction `QAlloc Mode,%reg` command can pre-allocate some cores for the requester (the pre-allocated cores appear for other cores as being in use, until either the requester core clears the preallocation or the requested core finishes calculation). This meta-instruction is compiled into the stream of executable code at compile time, but executed at runtime. Depending on actual HW availability, SV can or cannot provide the requested number of cores, i.e. the program must be prepared for both answers, using an `if`…`then`..`else` structure. Meta-instructions `QCreateT` and `QCreateF` create QTs in a way similar to `QCreate`, except that they do so only if the last `QAlloc` was successful and was not successful, respectively. If the best calculational facility is not available at the time of invocation, within the `QCreateF` QT, another, utilizing less resource-hungry allocation method can be attempted, to arbitrary depth (the recursivity is automatically provided by the physically different cores). This method enables the compiler to discover all available possibilities for performance-increasing parallelization and inserting hints about the executable code, while the processor can choose the maximum available option depending on actual HW availability.

*4.5. Handling subroutine call*

It can be noticed that a QT is very much similar to a subroutine: it has well-defined functionality, lifetime, address, changes the control to another place and after termination, returns to the instruction next to the place of calling; it returns a result. Its extra functionality is that that it receives another processing unit for performing its task. This provides extra possibilities, which can be utilized through using meta-instruction `QCallP`. This not only enables to organize the code in a modular way, but also enables the subroutine to work (at least partly) in parallel with the main code (`QWait` can be used to synchronize it with the main thread).

More important is, that EMPA QT subroutines have no duty to save/restore the calling address: this task is taken over by the control unit, without needing to store the return address. In this way the HW will not use the stack memory (i.e. no HW-placed items will interlace SW-placed items), which reduces memory traffic and simplifies parameter passing; it actually implements a kind of parallel inline call. In addition, cache will not be cooled down and no register save/restore needed: the subroutine also receives a new processing unit with its own register file and cache. Using EMPA-style subroutine calls considerably reduces both number of executed instructions and data traffic between processor and main memory.

*4.6. Coordinated operating mode*

One major problem with SPA in many-processor systems that they are segregated: other hardwares, typically shared memory, must eliminate the effects of the "random" operation of the cores. EMPA, however, can implement a *coordinated operating regime*. The EMPA way of implementing Direct Memory Access (DMA) is a nice example of the different thinking.

In SPA, CPU does not want to deal with every single byte during the I/O operation, so it enables the DMA device to access the memory directly. However, the computer is built from SPA components: CPU must share its data and address buses with the DMA device, which slows down both units. Notice that these two units work akin a restricted coordinated dual-issue processor: the CPU transfers data to DMA device and DMA device delivers termination signals to the CPU, both as a kind of inter-core communication.

In SPA, one can notice that the work of the two "processors" are fully coordinated. The CPU cannot expect a reasonable content in its buffer area while the transfer runs, and the DMA device is expected to access only its dedicated buffer part of the memory. Since in this way their operation cannot interfere with each other during transferring the data, they could work really in parallel, provided that not only the two processors have independent address and data handling capability, but also the memory (like [12]) and there are two independent buses available for transferring data in parallel. Both transfers can run at their full speed; and think about heavily loaded network routers or just a user downloading a picture from the network to its screen buffering data on its disk. The main obstacle here is SPA: all components designed with SPA in mind must be changed simultaneously; changing only one of them has no sense at all.

*4.7. Handling critical sections*

Sometimes (like when accessing some resource) it is critical to make it sure that some sequence of instructions will not be interrupted during execution. Since applications are not able to control directly the operation of the OS, in SPA an indirect way is chosen: using some OS services the application switches on an off a special protection. Because of context switching needed by the operation, this method is wasting time, memory cycles and code storage. In EMPA, a core can rent another core to execute the code in the critical section, and can wait until rented core finishes executing the critical section, uninterrupted. Again, the execution in the critical section can be run (at least partly) in parallel with the main code, at the price of utilizing one more core for the time of executing the critical section.

*4.8. Handling exceptions*

Exception handling is increasingly popular for implementing services in OSs for comfort and safety. The growing functionality and complexity of OS services also increased the execution time of such services, and especially in multiprocessing environment the frequency of their utilization. Essentially, this is why exception-related execution time does not get quicker proportionally with speeding up the HW [21].

In the today's environment one can enable luxury of dedicating one (or more, if needed) of the available cores for servicing exceptional control flow requests. Cores are properly initialized and after arriving to a `QWaitI` meta-instruction, they are waiting in kernel mode for receiving a request, possibly in power economy mode [31]. A user process can prepare parameter passing in user mode; the call to exception can pass the contents of the register file to those dedicated cores (waiting in kernel mode) using the method of core spilling [33]. In this way the context change is achieved through changing to a new core, rather than making lengthy processing and a lot of data movement; and even the kernel-core can run (at least partly) in parallel with the user-core, if it is properly organized and dependencies enable to do so. This feature essentially represents a user-space exception handling, with the safety and comfort of kernel-space servicing.

*4.9. Task scheduling*

The suggested operating regime represents a mapping between *n* processing units and *m* tasks. In SPA, the *m* tasks are mapped logically to the *same* processor (i.e. *m:1* mapping), even when the scheduling physically delegates several runnable tasks to several cores. In EMPA some (critical) tasks can be mapped to a dedicated processor and run permanently, some other tasks can be mapped to physically different processing units, even one task to several processing units, and also their execution times can overlap.

From the point of view of the OS, scheduling faces new possibilities and challenges. Some processes (like critical interrupt service routines and OS services) can be permanently scheduled (despite that they may be "blocked" in SPA sense) and most of the cores of the EMPA processor behave as a kind of "cache" for some processes: runnable processes can use computing resources in parallel, they can be (un)blocked without using OS services and memory cycles.

Task scheduling can be organized in a way similar to that of virtual pages. For the excessive processes conventional scheduling remains valid, but will only be activated when processor is really out of computing resources. This means that a computing resource bound task can run until it terminates, without being interrupted or slowed down due to the changing load of the computing system or seeing any "noise" of the OS; even when it utilizes OS services extensively. The consequence of this scheduling method are extremely low interrupt and OS service latency times and extremely low non-payload instruction execution ratio. Because the parallel execution of short QTs provides available cores in short times, this scheduling policy automatically prevents failures due to resource unavailability (like priority inversion), and so makes obsolete the special protocols used in SPA to handle those issues, and finally makes the OSs more effective and simple, comprising much less non-payload code.

*4.10. Taxonomy of the unconventional architecture*

Term RC covers essentially everything from ASICs to microprocessors [34] which at least partly utilizes unconventional principles and methods. The EMPA architecture suggested above does not fit the taxonomy [33]. The primary reason is, that all architectures in the taxonomy are built in SPA, i.e. are based on *segregated* processors, while EMPA uses *cooperating* processors. There is no question that EMPA has a place in category MCP, and similarly that its architecture is neither fixed nor hybrid. It is an open question, however, if it fits the branch 'Reconfigurable Architectures' or a newly to be introduced branch 'Configurable Architectures'. EMPA is RC in the sense that some fragments of the architecture are working in a regime that is typical in RC systems, including components (like block-RAM-like storages between the cores, extra non-conventionally working blocks outside the cores, dedicated wiring to implement inter-core communication) and operations (like cloning contents as bit arrays rather than register to register operations, or handling pseudo-registers). On the other hand, it is not RC in the sense that cores mostly work in conventional regime and routing takes place neither at the beginning nor when changing the architecture and the architecture is programmed (or better: configured) using conventional means.

*4.11. Other implications provoked by EMPA*

Recent HW developments demonstrated that even relatively small HW changes must be accompanied by corresponding changes in other levels of the computing stack. The case is not different for EMPA: everything needs (mostly transparent) changes, from electronic technology to compilation methods, but first of all: in thinking.

Theoretically, the basic paradigm did not change: there is an one-to-one correspondence between processor and process, and the timely behavior formerly seen in OSs is now implemented at processor level. Since QTs are formed by the compiler considering all kinds of dependencies, this extended computing paradigm is as good as the conventional one.

Engineers are thinking in inflexible architectures and predefined (mostly master-slave) hierarchy, which enable to use only one generation of linked cores, see for example [9]. In EMPA, this has been changed to a flexible, dynamic hierarchy, based on master-slave relationship, allowing several generations to work together.

**5. Developments for EMPA**

As the first step of implementing an EMPA processor a development system has been prepared [36]. The development started from the educational-purpose processor Y86 [20], because it models a widely used architecture (Intel x86), not overcrowded with SPA accelerators, and mainly it provides an easy path to implement an instruction group of different nature (meta-instructions) and a good starting point to prepare an open source development environment. An assembler being able to compile the needed mixture of executable and control codes and a digital simulator has been prepared. The simulator operates in hybrid mode: the executable instructions are simulated using the (slightly modified) ISA level simulator of Y86, while the EMPA-related operations are executed in a cycle-accurate mode in a C++ simulator.

*5.1. Tools for studying EMPA operation*

As it might be guessed, neither to implement nor to understand the details of operation of EMPA is not simple. The simulator is implemented in two versions. The one with graphic GUI uses Qt [37] library to provide visual insight into the complex operation. The command-line based simulator prepares extensive logging, available for further processing. Both versions prepare a processing diagram, depicting graphically the rather sophisticated internal operation.

The processing diagram (for an example see Fig. 5) is a by-product of the cycle-accurate simulator and attempts to visualize the rather complex internal operation of the EMPA processor. The diagram should indicate, at which time, by which core, which instruction was executed; how the cores interacted with one another; and whether cores execute conventional executable or meta-instructions. Thus, a lot of information must be crowded into the figure.

A processing diagram shows cores on the horizontal axis and time on the vertical axis. For better orientation, grid lines are placed at every 5th clock cycle. The length of a clock cycle is the length of a control operation, the instruction execution is supposed to be of variable length. Arbitrary, but reasonable instruction lengths are assumed.

Rectangular blocks represent QTs, with hooks at their top and bottom, for their creation and termination, respectively. In columns $C_x$ the vertical rectangles represent the "lifetime" of a QT. At the times outside of the QT rectangles, the core is in power economy mode, not running a QT.

The parent-child relationship is illustrated with labels of the QTs: the first few chars are identical with those of the parent, and the last char denotes the sequence number of the child. For the human reader, in the figure (as well as in the corresponding simulator log files) core sequence numbers and textual QT ID strings are shown rather than the "one-hot" bitmasks used internally by the simulator.

The memory address of a meta-instruction is shown on the right side of the QT in a square box, the address of an executable instruction is shown on top of a bigger ball, and some smaller balls represent the duration of the instruction. While a core is waiting, at the corresponding time a circle with the respective memory address is displayed at the left side of the QT block. From memory address the corresponding source code line can be found using the program listing. For a more detailed legend and further processing diagrams see [36].

Fig. 5 shows processing diagram of the direct implementation of the parallelism shown in Fig. 1, on the left side when the processor's control unit is able to provide all the needed 4 cores. On the right side the case when there are not enough cores is shown: if the processor has only 3 free cores for solving the task, the QT in question gets blocked until the needed core is put back in the pool.

This latter example also demonstrates how EMPA can solve the problem of mapping the virtually infinite number of processes to the finite number of available cores: the QTs which do not find available computing resource must wait (and so the issuing QT gets blocked until a "reprocessed" processing unit gets available). The execution time is slightly longer, the requesting core must wait. The compiled code, however, is the same: the compiler could discover all parallelization possibilities, but actual HW availability enabled to utilize part of them: the code adapts to the actual HW availability. The actual core allocation policy is of course more sophisticated, for details see [36].

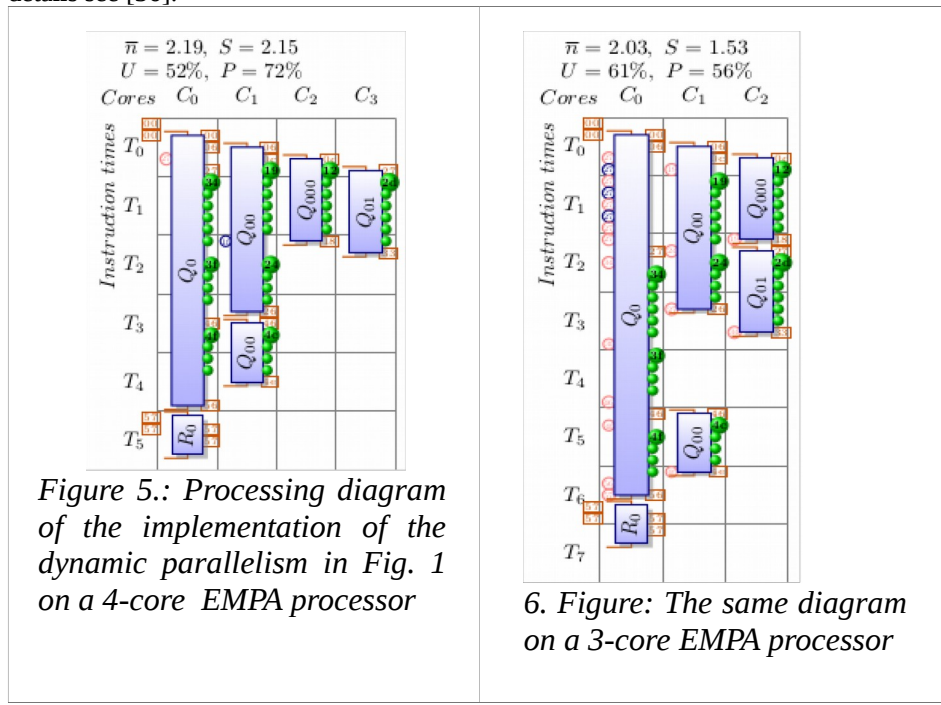

*Figure 5.: Processing diagram of the implementation of the dynamic parallelism in Fig. 1 on a 4-core EMPA processor*

*6. Figure: The same diagram on a 3-core EMPA processor*

*5.2. Imitating conventional HW parallelization solutions*

For different calculation tasks, a wide variety of parallelization solutions has been elaborated [2], and works excellently for the calculation they were designed for. For another type of calculations, they are not useful, unnecessarily increase complexity and power consumption. A reasonable expectation against a dynamic architecture is to be able to imitate those useful structures, if the compiler recognizes their utilization would be advantageous. Fortunately, the known such constructions can be composed of the components dynamically, see examples in [36]. As an example, the code constructing dynamically an architecture for implementing a speculative evaluation

```
C = B-A > C ? D+3 : E+4;
```
is as follows:

```
0x000:                  |         .pos 0   # Program starts at address 0000
0x000: f5f142000000 | QTMainC: QCreate QTMainT,%ecx #Create wrapper QT
0x006: 500f6c000000 |      mrmovl A,%eax   # Load the variable operand
0x00c: f844000000   |      QCallP  QTposC  #  Result if >0
0x011: f857000000   |      QCallP  QTnegC  #  Result if <= 0
                    |      # Make the main calculation while speculating
0x016: 500f6c000000 |      mrmovl A, %eax  # Load variable operand
0x01c: 503f70000000 |      mrmovl B, %ebx  # Load fix operand
0x022: 6103         |      subl   %eax,%ebx #
0x024: 7638000000   |      jg     Plus
0x029: f044000000   |       QWait   QTposC # Load latched bad result
0x02e: f057000000   |       QWait   QTnegC  # Load latched good result
0x033: 7042000000   |       jmp    Ready
0x038: f057000000   | Plus: QWait   QTnegC  # Load latched bad result
0x03d: f044000000   |       QWait   QTposC  # Load latched good result
0x042:              | Ready:
0x042: f3           | QTMainT: QTerm    # Return the last core!!!
0x043: 00           |        halt
                    | # The "subroutine QTs" used
                    | # These are computed in advance, the speculation
0x044: f5f156000000 | QTposC:  QCreate QTposT, %ecx # Computation
0x04a: 501f74000000 |           mrmovl D,%ecx
0x050: c0f103000000 |           iaddl  3,%ecx
0x056: f3           | QTposT:  QTerm    # Return operand1 in %ecx
0x057: f5f169000000 | QTnegC:  QCreate QTnegT, %ecx # Computation
0x05d: 501f78000000 |           mrmovl E,%ecx
0x063: c0f104000000 |           iaddl  4,%ecx
0x069: f3           | QTnegT:  QTerm    # Return operand1 in %ecx
```

Listing 1: Coding to assemble an ad-hoc computing setup for speculative evaluation

The main calculation is to load A and B, and perform subtraction. The speculation is started before starting the main calculation: both branches are prepared and started, and the speculative calculations run in parallel with the main calculation, on newly rent cores. When the main calculation finishes, the condition flags are set properly, and a conditional jump selects one of two branches, where the 'correct' results can be read and the 'wrong' results must be discarded.

Since both results are present in latch registers of the parent, reading of results of both branches must be triggered by a specific 'wait' instruction. Since the latches for parent core stores also the offset of the QT which provided the back-linked register value, both branches can safely return their result in register %ecx: they are different, because they are prepended by the offset value of the QT (of course the results must be returned by both branches in the same link register).

The trick the program uses is that *two* waits are used on both branches. The first one (on both branches) reads the 'wrong' value (from the loser QT) into register %ecx of the parent. Immediately after this, the 'correct' value (from the winner QT) is read also in register `%ecx`, and so it overwrites the wrong value. In this way the joint execution that follows will use the correct value, which becomes known *after* that both speculative evaluations terminated. This ad-hoc architecture reproduces the functionality of speculative evaluation, but does not use inflexible resources.

```
0x015: 506100000000 | Loop:     mrmovl (%ecx),%esi # get *Start
0x01b: 6060         |     addl %esi,%eax    # add to sum
0x01d: 30f304000000 |     irmovl $4,%ebx
0x023: 6031         |     addl %ebx,%ecx    # Start++
0x025: 30f3ffffffff |     irmovl $-1,%ebx
0x02b: 6032         |     addl %ebx,%edx    # Count--
0x02d: 7415000000   |     jne  Loop         # Stop when 0
```

Listing 2: The conventional coding for the vector sum-up task

*5.3. Computing performance*

The coding of the example task mentioned, summing up elements of a vector, is formulated for three versions in this subsection. As mentioned, the conventional coding (see listing above) comprises 7 instructions, of which only one is a payload instruction.

In EMPA (see listing below), a special looping method helps to implement that task. In the first line, a helper core is allocated. The joint operating mode is **Mode1**, which is set in both partners. In **Mode1**, the control unit will take over loop control and repeats creating the QT according to the value found in register `%edx`. On a new iteration the pointer to the actual array element is increased by the length of the element and the cycle counter is decreased between creating the adjacent QTs. The content of the address is cleared (can be used as offset) by `QAlloc` and is advanced to the beginning of the array before looping to provide address. The sum is formed in register `%eax` (clearing not displayed).

```
0x00e: f4f201        |    QAlloc Mode1, %edx # allocate %edx times
0x011: 201d          |      rrmovl %ecx, %esv#Write array address
0x013: f6f021000000  | QT1LoopC:QCreate QT1LoopT, %eax
0x019: 501d00000000  |      mrmovl (%esv), %ecx # The summand
0x01f: 6010          |      addl   %ecx, %eax # Add it to sum
0x021: f0            | QT1LoopT:QTerm
```

Listing 3: The coding for the vector sum-up task using the FOR method of EMPA

This code is executed jointly by the two cores. Executing meta-instruction `QCreate` in parent core rents one child core (the one preallocated by `QAlloc`), tells the SV the address of the array. The child QT receives the contents of all registers as well as the address of the actual array element from the control unit and executes the body: picks up the summand and adds it to register **%eax**. Notice how the pseudo-register works: the parent writes the address into 'its own' register **%esv**, and the actual child reads the actual address from 'its own' register **%esv**. Actually both of them talk to SV, and SV is allowed to make any operation on the (apparent) content: advance the pointer and decrease cycle counter (the same way as content of stack pointer changes in SPA).

Executing **QTerm** terminates child QT and returns in link register **%eax** the updated partial sum, which will replace content of **%eax** in the parent core. The child core is returned to the pool, but remains pre-allocated. The **QTerm** in the child causes unblocking the parent QT. Since SV kept the PC in the parent at the address of **QCreate** and the core remains pre-allocated, the process will be repeated until the maximum iteration count reached. However, the working core is not available for the next iteration until the previous iteration finishes, so the parent QT must wait for it (i.e. it gets blocked). After the iteration count reached, the SV considers the QT as executed, clears the pre-allocation and PC in parent jumps to the instruction next to **QTerm**.

The method above is moderately resource-hungry: it uses two cores and executes only 2 instructions instead of 7 instructions used in the traditional programming. Notice that the meta-instructions are extremely simple control instructions which can be 'executed' using a much higher clock frequency, and that the two executable instructions (rather than one) are needed only because the instruction set of Y86 cannot make the operation in one single step.

Another approach can be to use as many cores as many summands we have. EMPA defines another cooperation regime for this mode, see Listing 4. The main difference is that in this mode a core is reserved for each summand, rather than only one for all calculations. The obstacle here is that the second core must wait termination of the first summing, i.e. the performance advantage is lost, because nearly all cores must nearly all time wait. The EMPA architecture, however, provides a solution to this problem and enables to parallelize this classically non-parallelizable task.

```
0x00e: f4f205        |        QAlloc Mode5, %edx  #Preallocate %edx cores
0x011: 201d          |        rrmovl %ecx, %esv      #Write array address
0x013: f6ff21000000  | QTLoopC:QCreate QTLoopT, %eno #%esv sums up
0x019: 501d00000000|         mrmovl (%esv),%ecx #get *Start+Index
0x01f: 601d          |        addl %ecx, %esv # Sum in parent's %esv
0x021: f0            | QTLoopT:QTerm
0x022: f113000000    |        QWait   QTLoopC # Wait until child ready
0x027: 20d0          |        rrmovl %esv, %eax      # Make result visible
```

Listing 4: The coding for the vector sum-up task using the SUMUP method of EMPA

In **Mode5** the controller pre-allocates the requested number of cores and provides an internal adder for the task. The PC of the parent core will stay pointing to **QCreate**, but unlike in the previous case there are available cores for the operation, so in adjacent clock cycles the parent can create further QTs. Since the actual array address is advanced between creating QTs, the adjacent QTs will deal with another elements. Another difference that the child QT does not return any result, rather it makes the summing in register `%esv`. As discussed, this register is controlled by the control unit, so in this mode the summand is transmitted to the adder in the parent, which sums up in the provided adder the summands received in consecutive cycles.

As the loop counter decreases to zero, the PC of the parent advances to the instruction next to `QTerm`. It is very probable, however, that at that time some of the children are still working. The `QWait` instruction makes sure that all summands arrived, and then the parent makes the sum visible: copies `%esv` to `%eax`; in this mode this is the functionality. Notice that in this mode *writing* `%esv` sends array address to the future child, and *reading* `%esv` reads the final sum from the specially allocated adder (much similar to the behavior of a control/status register).

*5.4. Performance of the different modes of summing*

For the sake of simplicity, neither of the above examples considered whether the computing resource allocation was successful. As mentioned above, the code can adapt itself to the actual HW availability conditions. The adaptive code first attempts to pre-allocate as many cores as many items are in the summing. If this resource-hungry pre-allocation method fails, it attempts to pre-allocate one core to utilize at least the FOR mode. If this fails, too, than only the conventional method remains: just *compute* the values/signals which can be provided by partners in an EMPA architecture. Anyhow, the adaptive code will deliver the resulting sum, but of course, the performance strongly depends on which method was enabled by the actual HW availability.

The simulator [36] enables to evaluate the performance in a quantitative way. The simulator utilizes arbitrary (but reasonable) instruction execution times, expressed in units of control clock cycles. For the three implementations, the execution time `T` depends on the vector length `n` as

```
T = 22 + n*30    //NO
T = 20 + n*11    //FOR
T = 32 + n*1     //SUMUP
```

Based on these slopes, FOR mode of EMPA is nearly *3 times quicker* than the conventional method, as some computed control statements are replaced by the much more effective control functionality. This requires only 2 cores.

In SUMUP method, in addition to omitting computed control machine instructions, even the obsolete fetch, decode, write-back, etc. stages of one instruction execution in the loop kernel are replaced by control functionality. In our setup the SUMUP method is *30 times quicker* than the conventional (NO) method, at the cost of using 30 more additional cores. This behavior is especially valuable, because *the algorithm cannot be parallelized at all using conventional methods of parallelization*.

Measured speedup values are derived from a mixture of different types of instructions: both conventional and EMPA codes contain both sequential and parallel parts, so despite the linear increase, the measured speedup will not linearly depend on the vector length, see Fig. 7, left side. The two speedup values will saturate for large vector lengths at values 30/11 and 30, respectively.

*5.5. Efficiency of parallelization*

Because EMPA omits (i.e. replaces with much quicker control facilities) some machine instructions, the relative speedup can even exceed unity, see Fig. 7, right side. This *does not not mean* a higher processor performance, it is due to the more clever organization of loops (less machine instructions).

In SUMUP mode, helper cores are only utilized for a short period of time, so the utilization efficiency [29] is low for short vectors. The returned PUs, however, are immediately available for other calculations in a more complex case.

Note that since cores are put back in the pool, much lower number of cores may be needed for very long vectors. If the compiler can find out the length of processing in that mode (in our setup it is 30 clock cycles), it should allocate not more than that number of cores: when the parent needs to add the 31st element, the 1st core is available again, so summing can be continued for an arbitrary vector length, needing only 30 cores at a time, and providing speedup 30.

For SUMUP method, the two different points of view of the two merits, the widely used *efficiency* and the recently introduced *effective parallelism* [16], are nicely reflected on the figure. For short vectors, *effective parallelism* is relatively low, because only fragment of helper cores are used only in a fragment of time. As all the 30 helper cores have a "full time job", the *effective parallelism* dependence saturates at value 1. In contrast, *efficiency* starts to decrease with increasing number of cores, and after reaching 30 cores, the speedup continues, but the number of cores involved in the calculation remains constant, so the dependence turns back and also saturates at value 1, but approaches it much more slowly.

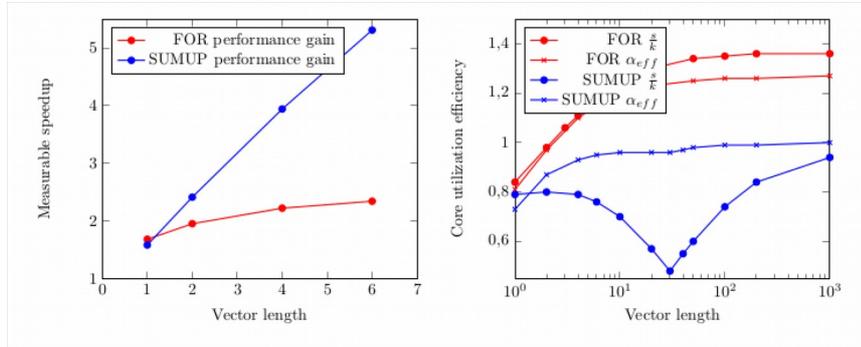

*Figure 7: The measurable speedup (left side) and the core utilization efficiency (right side) for two different mass processing methods, in function of the vector length.*

*5.6. Handling exceptions*

A useful way of cooperation is when a core is reserved (and configured) to service OS requests or asynchronous interrupts. The corresponding QT is created in advance and waits in supervisor mode for an activation signal from the SV. Because upon service request the child core inherits context of the requesting core, it can continue the operation in supervisor mode *without needing a context change*: neither processor nor process context must be saved. In systems heavily loaded with asynchronous interrupts or using a lot of OS services, one-two orders of magnitude higher execution speed can be achieved in these spots. Furthermore, kernel code can run (at least partly) in parallel with parent QT, running in user mode (in another core). A special advantage of EMPA is that the reserved core has its own "hot" in-core cache memory which is not cooled down during servicing. Similarly, the in-core cache of the parent remains hot.

The speed advantage of using alternative implementation instead of the conventional one was proved experimentally, using a reconfigurable electronic model. The alternative electronic model, even after linking it to the conventional computing system, achieved a factor of 30 speed advantage [19].

**6. Summary**

The present crisis of computing made clear that the development of computing really cannot be continued using the 70-years old computing paradigms, completely neglecting both the requirements against computing and the achievements of the technological development. It looks like that the main obstacle is the too rigid interpretation of the computing paradigms. A relatively simple change (recognizing that a process can be mapped to one of the available processors, rather than the only available processor must be shared among several processes) opens new perspectives of computing. The suggested flexible on-demand type architecture suggests better ways of separating computing into layers, makes computer executing times predictable, using OS services less expensive in terms of execution time, enables to increase performance of single-threaded programs through recompiling rather than rewriting them.

Although the new approach requires changes in all fields and all levels of computing, the suggested changes are a kind of generalization of the former paradigms; i. e. the new computing is upward compatible with the old one. The suggested approach is also beneficial for solving the energy consumption and "dark silicon" issues: the unused cores can be kept in power economy mode, and the core availability may depend also on its temperature.

**Acknowledgement** Project no. 125547 has been implemented with the support provided from the National Research, Development and Innovation Fund of Hungary, financed under the K funding scheme.

**References**

[1] S. H. Fuller and L. I. Millett, *The Future of Computing Performance: Game Over or Next Level?*, National Academies Press, Washington, 2011.
[2] K. Hwang and N Jotwani, Advanced Computer Architecture: Parallelism, Scalability, Programmability, Mc Graw Hill, 2016,
[3] K. Asanovic et al, A View of the Parallel Computing Landscape, Comm. ACM, **52**(2009)56-67.
[4] M. S. Schlansker, and B. R. Rau, EPIC: Explicitly Parallel Instruction Computing, Computer **33(**2000)37–45.
[5] M. D. Hill and M. R. Marty, Amdahl's Law in the Multicore Era, IEEE Computer 41(2008)33-38.
[6] U. Vishkin, Is Multicore Hardware for General-Purpose Parallel Processing Broken?, Comm. ACM **57**(2014)35.
[7] Esmaeilzadeh, Hadi, Approximate Acceleration: A Path Through the Era of Dark Silicon and Big Data, Proceedings of the 2015 International Conference on Compilers, Architecture and Synthesis for Embedded Systems (CASES '15) pp. 31–32, 2015.
[8] K. Pingali et al, The Tao of Parallelism in Algorithms, *SIGPLAN Not.*, **46**(2011)12–25
[9] U. Vishkin, Using Simple Abstraction to Reinvent Computing for Parallelism, Comm. ACM, **54**(2011)75-85
[10] W Aspray, John von Neumann and the Origins of Modern Computing, pp. 34--48 , MIT Press, Cambridge, 1990
[11] T. David, R. Guerraoui, and V. Trigonakis, Everything you always wanted to know about synchronization but were afraid to ask, Proc. of the Twenty-Fourth ACM Symposium on Operating Systems Principles (SOSP '13), pp 33-48, 2013.
[12] Cypress,CY7C026A: 16K x 16 Dual-Port Static RAM, http://www.cypress.com/documentation/datasheets/cy7c026a-16k-x-16-dual-port-static-ram, 2015.
[13] V. W. Lee at al, Debunking the 100X GPU vs. CPU Myth: An Evaluation of Throughput Computing on CPU and GPU. Proce 37th Ann. Internat. Symp. on Computer Architecture, 451-460, 2010.
[14] G. M. Amdahl, Validity of the Single Processor Approach to Achieving Large-Scale Computing Capabilities, AFIPS Conf. Proc.s **30**(1967) , 483-485
[15] TOP500.org, The Top 500 supercomputers, (2016) https://www.top500.org
[16] János Végh and Péter Molnár, How to measure perfectness of parallelization in hardware/software systems (2017) 18[th] Internat. Carpathian Control Conf. (ICCC) Paper 121
[17] P. J. Denning and T. G. Lewis, Exponential Laws of Computing Growth, Comm. ACM **60**(2017) 54-65.

[18] R. Hameed et al, Understanding Sources of Inefficiency in General-purpose Chips, Proceedings of the 37th Annual International Symposium on Computer Architecture (ISCA '10), pp 37–47, **2010**.

[19] J. Végh et al, An alternative implementation for accelerating some functions of operating system, ICSOFT-EA 2014 – Proc. 9th Internat. Conf. on Software Engineering and Applications, pp 494-499, 2014.

[20] R. E. Bryant, and D. R. O'Hallaron, Computer Systems: A Programmer's Perspective, Pearson, 2014.

[21] D. Tsafrir, The Context-switch Overhead Inflicted by Hardware Interrupts (and the Enigma of Do-nothing Loops). *Proc. 207 Workshop of Experimental Computer Science*, art. No 4, 2007.

[22] Arvind and R. A. Iannucci, *Two Fundamental Issues in Multiprocessing*, 4th International DFVLR Seminar on Foundations of Engineering Sciences on Parallel Computing in Science and Engineering, pp. 61--88 , 1988.

[23] T. Ungerer et al., MERASA: Multi-core execution of hard real-time applications supporting analyzability, *IEEE Micro* **99**(2010)66-75.

[24] J. Backus, Can Programming Languages Be Liberated from the von Neumann Style? A Functional Style and its Algebra of Programs, *Comm ACM*, **21**, 1978, 613-641.

[25] A. Nicolau and J. A. Fisher, *Measuring the parallelism available for very long instruction word architectures*, IEEE Transactions on Computers **C-33**(1984) 968-976.

[26] D. W. Wall, Limits of Instruction-Level Parallelism, 1993. http://www.hpl.hp.com/techreports/Compaq-DEC/WRL-93-6.pdf

[27] S.A. Mahlke et al, Scalar program performance on multiple-instruction-issue processors with a limited number of registers, Proc. Twenty-Fifth Hawaii International Conference on System Sciences, PP. 34 – 44, 1992.

[28] J. K. Ousterhout, Why Aren't Operating Systems Getting Faster As Fast As Hardware?, USENIX Summer Conference, 1990.

[29] J. Végh, A configurable accelerator for manycores: the Explicitly Many-Processor Approach, ArXiv e-prints 1607.01643, 2016.

[30] J. Congy et al, IEEE Trans. Accelerating Sequential Applications on CMPs Using Core Spilling, *Parallel and Distributed Systems*, **18**(2007)1094 - 1107

[31] S. Borkar, *Proc. 44th Annual Design Automation Conference DAC'07*, Thousand Core Chips: A Technology Perspective, pp. 746-749, 2007.

[32] J. Williams et al, *Computational density of fixed and reconfigurable multi-core devices for application acceleration*, Proc. Reconf Systems Summer Institute, 2008

[33] J. Williams et al., Characterization of Fixed and Reconfigurable Multi-Core Devices for Application Acceleration, *ACM Trans. Reconfigurable Technol. Syst.* **3**(2010) 1936-7406

[34] K. Compton and S. Hauck, Reconfigurable Computing: A Survey of Systems and Software, *ACM Comput. Surv.* **34**(2002) 171–210.

[35] X. Wen and U. Vishkin, *FPGA-based prototype of a PRAM-on-chip processor*, Proc. 5th conf. on Computing frontiers, CF'08, 2008.

[36] J. Végh, *EMPAthY86: A cycle accurate simulator for Explicitly Many-Processor Approach (EMPA) computer*. 10.5281/zenodo.58063, https://github.com/jvegh/EMPAthY86 2016.

[37] J. Blanchette and M. Summerfield, A Brief History of Qt. C++ GUI Programming with Qt 4. Prentice-Hall, 2006.